\documentclass{myeas}
\usepackage{graphicx}
\DeclareRobustCommand{\ion}[2]{%
\relax\ifmmode
\ifx\testbx\f@series
{\mathbf{#1\,\mathsc{#2}}}\else
{\mathrm{#1\,\mathsc{#2}}}\fi
\else\textup{#1\,{\mdseries\textsc{#2}}}%
\fi}
\def\la{\mathrel{\mathchoice {\vcenter{\offinterlineskip\halign{\hfil
$\displaystyle##$\hfil\cr<\cr\sim\cr}}}
{\vcenter{\offinterlineskip\halign{\hfil$\textstyle##$\hfil\cr
<\cr\sim\cr}}}
{\vcenter{\offinterlineskip\halign{\hfil$\scriptstyle##$\hfil\cr
<\cr\sim\cr}}}
{\vcenter{\offinterlineskip\halign{\hfil$\scriptscriptstyle##$\hfil\cr
<\cr\sim\cr}}}}}

\def\aj{AJ}%
\def\araa{ARA\&A}%
\def\apj{ApJ}%
\def\apjl{ApJ}%
\def\apjs{ApJS}%
\def\aap{A\&A}%
\def\mnras{MNRAS}%
\def\pasp{PASP}%
\def\nat{Nature}%
\begin{document}

%%-----------------------------
%%      the top matter
%%-----------------------------
\title{Low metallicity stars in our Galaxy
} 
\author{Piercarlo Bonifacio}\address{
CIFIST Marie Curie Excellence Team
--
Observatoire de Paris, GEPI,
             F-92195 Meudon Cedex, France}
\secondaddress{
Istituto Nazionale di Astrofisica - Osservatorio Astronomico di
Trieste,
    Via Tiepolo 11, I-34131
}
%\author{...}\address{...}
%\author{...}\address{...}
%
%
\begin{abstract}
The advent of 8\, m class telescopes has allowed the detailed spectroscopic
study of sizeable numbers of extremely metal-poor Galactic stars which
are the witnesses of the formation of the early Galaxy.  Their chemical
composition displays some distinctive trends which should  provide
a strong constraint on the physical nature of the first generation(s) of
stars and on their nucleosynthetic output. I will review  recent results 
in the field following the periodic table, from lithium to uranium and
shortly comment on  the intriguing classes of Carbon Enhanced Metal Poor (CEMP) 
stars, for many of which there is no analogue among solar metallicity  stars. 
In spite of these exciting results, the number of known stars of metallicity 
below [Fe/H]=--3.3 remains quite small and it would be desirable to discover 
more, both to clearly understand the metal-weak tail of Halo metallicity 
distribution and to clarify the abundance trends at the lowest metallicities. 
Most of these extremely rare objects have been discovered 
by the wide field objective prism  surveys,  HK survey and  Hamburg-ESO survey.
In the near future the Sloan Digital Sky Survey and its continuation SEGUE 
are expected to boost significantly the numbers of known extremely metal poor 
stars.  We are living exciting times but an even more exciting future
lies ahead !
\end{abstract}
\maketitle
%%-----------------------------
%%      your text
%%-----------------------------
\section{Introduction}

In this review I will try to outline  the observational
picture of the chemical composition of extremely metal-poor stars
as it emerges from  recent investigations.
I will not dwell on interpretations of the chemical patterns,
since this topic will be covered by other speakers at this conference.

\section{Lithium}

Lithium is the heaviest of the nuclei produced primordially
in the first three minutes of the existence of the Universe
and the lightest of the ``metals'', in the astronomical acception
of the term.
Since, over 20 years ago, Monique and Fran\c cois Spite
discovered that warm metal-poor stars share the same
lithium abundance (the so-called {\em Spite plateau}),
and interpreted this as the lithium
primordially produced
(Spite \& Spite \cite{spite82a,spite82b}), a very active 
research has been carried out in the field. 
The major challenge to the interpretation of the
{\em Spite plateau} as evidence for a primordial production
of lithium comes from the value of $\eta$ 
(the baryon to photon ratio) inferred from the fluctuations
of the Cosmic Microwave Background, as measured by the
WMAP satellite (Spergel et al. \cite{spergel,spergel06}).
When used in Standard Big Bang Nucleosynthesis
computations, this value implies 
A(Li)\footnote{here and after, for any element X
we define the logarithmic abundance as
A(X) = log[N(X)/N(H)] +12.}=2.64.
The highest values claimed for
the {\em Spite plateau} 
(Bonifacio et al. \cite{B02};
Mel{\' e}ndez \& Ram{\'{\i}}rez \cite{melendez} 
are about 0.3 dex lower; many other
recent claims are even lower (see Bonifacio et al.
\cite{B06}, and references therein).
This  has also been called the ``cosmological
lithium discrepancy'' (Korn et al. \cite{Korn}). 

To complete the observational picture 
one should add that 
some investigations suggest a slope 
in  Li abundance vs. [Fe/H]\footnote{
For any two elements X,Y, the notation [X/Y] stands for
log(X/Y)-log(X/Y)$_\odot$,
where $\odot$ indicates the solar value
and X and Y indicate the number of atoms 
of species X and Y, respectively.
}
of about  0.1--0.2 dex/dex 
(Ryan et al. \cite{ryan96,ryan};
Boesgaard et al. \cite{boesli};
Asplund et al. \cite{asplund06}),
while other investigations fail to detect
any slope
(Spite et al. \cite{spite96}; Bonifacio \& Molaro
\cite{BM97}; Bonifacio et al. \cite{B02};
Mel{\' e}ndez \& Ram{\'{\i}}rez \cite{melendez},
Bonifacio et al. \cite{B06})
or find a very shallow one
(Charbonnel \& Primas \cite{CP05}).
The presence of such a slope clearly does not allow
to interpret  
the {\em Spite plateau} as primordial lithium.

Finally an exciting result is the measurement
of $^6$Li in metal-poor stars, although
the interpretation of these measurements
is by no means straightforward.
A non-controversial measurement of
$^6$Li was achieved 
in the 1990's 
for only one halo star: HD~84937 
(Smith et al. \cite{smith93};
Hobbs \& Thorburn \cite{HT1994};
Smith et al. \cite{smith98};
Cayrel et al. \cite{cayrel99}),
with 
 $^6$Li/$^7$Li $\sim  5$\%.
Subsequently, measurements of  
$^6$Li have been claimed for three other halo stars, but
have been withdrawn on the basis of higher quality spectra.
Smith et al. (\cite{smith98}) measured
$^6$Li in BD $+26^\circ 3578$ (= HD~338529),
but Asplund et al. (\cite{asplund06})
changed this to an upper limit.  
Deliyannis \& Ryan (\cite{DR2000})
measured 
$^6$Li in HD~140283 but 
Aoki et al. (\cite{aoki})
changed this to an upper limit.
Nissen et al. (\cite{nissen00}) measured
$^6$Li in G~271-162, and also
for this star Asplund et al. (\cite{asplund06})
changed it to  an upper limit.
The investigation of Asplund et al. 
(\cite{asplund06}) lead to the measurement of $^6$Li in nine more halo stars.
All the 10 stars with $^6$Li measurements
have the same Li isotopic ratio 
 $^6$Li/$^7$Li $\sim $
5\%; 
we are therefore in presence of a 
$^6$Li plateau which  mirrors the $^7$Li plateau.

Korn et al. (\cite{Korn}) have measured Li 
and Fe in stars of different luminosity 
in the globular cluster NGC 6397 and found lower
Fe and Li abundances in the Turn-Off stars
than in the sub-giant stars. They argue that this
is a result of atomic diffusion in the atmospheres
of these stars and claim that this could ``solve''
the cosmological lithium problem.
This work is extremely suggestive, however
it should be noted that this result is at odds
with previous analysis of this same cluster
(Castilho et al. \cite{castilho}; Gratton et al.
\cite{gratton}), and relies in a crucial way
on the temperature scale adopted, in particular
on the effective temperature assigned to the TO stars.
An increase of 100 K of the assumed TO temperature 
would effectively erase the abundance differences found.

\section{Beryllium}

The only production channel for Be is 
cosmic ray spallation: either fast C, N and O nuclei
hit H and He in the interstellar medium (ISM), or fast
protons and $\alpha$ particles hit C, N or O nuclei
in the ISM. Whichever the case, light nuclei
$^6$Li, $^7$Li, $^9$Be, 
$^{10}$B and $^{11}$B are produced.
It has been known  for 
over 20 years that metal-poor
stars show low Be abundances
(Molaro \& Beckman \cite{mol84})
and that there is a linear
rise of Be abundance
with metallicity
(Gilmore et al. \cite{gilmore};
Ryan et al. \cite{ryan92};
Molaro et al. \cite{mol97};
Boesgaard et al. \cite{Boesgaard1999};
Boesgaard \& Novicki \cite{Boesgaard2006}
).
This well established trend, and the simplicity
of the Be production mechanism, suggested
its possible use as a chronometer
(Beers et al. \cite{beers2000};
Suzuki et al. \cite{suzuki}).
The measurement of Be in two Globular
Clusters, NGC 6397 (Pasquini et al. \cite{Pasq04})
and NGC 6752 (Pasquini et al. \cite{Pasq06}),
and the good agreement between the 
``Be ages'' and the ages derived from 
Main Sequence fitting, demonstrated the 
usefulness of Be as a chronometer.
Furthermore Pasquini et al. (\cite{Pasq05})
were able to show that, independently of
the absolute calibration of the ``Be age'',
Be abundances can be usefully used as
a ``time-like'' axis to study the evolution
of other abundance ratios, allowing to discriminate
the various Galactic populations (e.g. Halo vs. 
Thick Disc).

Beryllium measurements rely on the 313\, nm \ion{Be}{ii}
resonance doublet, quite near to the atmospheric
cut-off,  and are always observationally challenging.
Our ability to fully exploit the chronometric
opportunities of this element in the future, will depend 
on the existence of high resolution spectrographs,
efficient in the near UV,
on large or extremely large telescopes.

\section{Carbon and Nitrogen}

At the lowest metallicities our knowledge of these 
two elements, is mainly due
to giant stars in which molecular lines of CH, CN, NH
can be measured down to very low metallicities.
This poses immediately a problem:  the abundances
of these elements in the atmospheres of giants may
be altered due to mixing episodes.
Luckily both the Li abundances and the combined
C and N abundances can be used to discriminate
observationally between ``mixed'' and ``unmixed'' 
giants (Spite et al. \cite{spite05}).
In fact examination of the isotopic ratio
$^{12}$C/$^{13}$C,  allows to conclude that 
the mixing in luminous RGB stars is more 
extensive than predicted in ``standard'' models
(Spite et al. \cite{spite06}).

\begin{figure}
\includegraphics[width=\hsize]{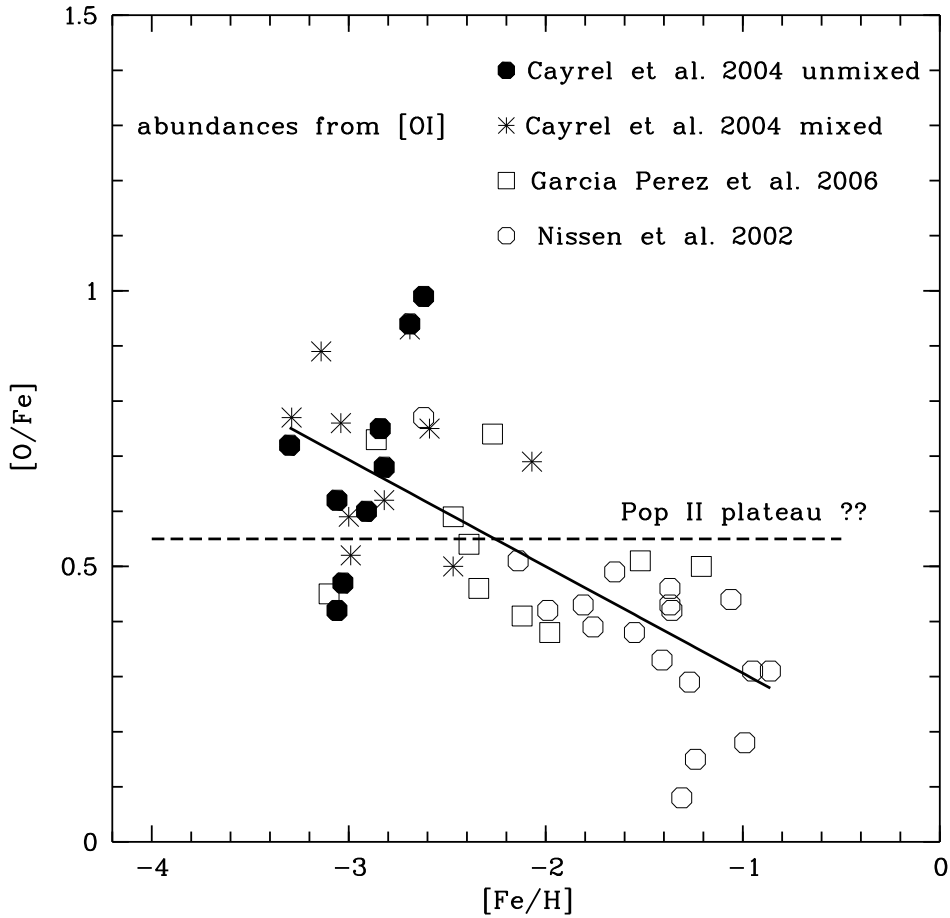}
\caption{Recent measurements of oxygen 
in metal poor stars, based {\em only}
on the [OI] 630\, nm forbidden line:
unmixed giants from Cayrel et al. (\cite{cayrel04},
filled hexagons), mixed giants from Cayrel et al.
(asterisks), subgiants from Garc{\'{\i}}a P{\'e}rez et al.
(\cite{ana}, open squares), main sequence
and subgiant stars from Nissen et al. (\cite{nissen02}).
The dotted line shows the Pop II plateau
claimed by  Garc{\'{\i}}a P{\'e}rez et al., while 
the solid line is a simple linear regression
to the available data. 
The [Fe/H] values of Nissen et al. and
Garc{\'{\i}}a P{\'e}rez et al. have been 
shifted by --0.2 dex, in order to be on the
same scale of the data of Cayrel et al.,
as described in Bonifacio et al. (\cite{B06}).
The [O/Fe] values, instead, have not been 
altered, since for all the authors they reflect
[O/\ion{Fe}{ii}], which is not overly
sensitive to errors in surface gravity. 
}
\end{figure}

The ``unmixed'' giants can be used as reliable
tracers of the Galactic chemical evolution.
Looking only at ``unmixed'' giants one finds
that that the [C/Fe] and [C/Mg] ratios are constant
with metallicity, or in other words, carbon
tracks iron and magnesium.
Also for nitrogen one finds that
[N/Fe] and [N/Mg] are  constant with metallicity,
at least for  [Mg/H] $\le -1.4$,
at higher metallicities the [N/Mg] rises from 
the value of $\sim -0.8$ up to 
the solar value (Spite et al. \cite{spite05};
Israelian et al. \cite{israelian04}).
The existence of this nitrogen 
``plateau'' at  low metallicities,
implies that nitrogen cannot be treated as
a pure ``secondary'' element.
At extremely low metallicities
Spite et al. (\cite{spite05})
have argued that there could be 
a second ``plateau'' at [N/Mg]$\sim -1.4$,
reminiscent of what has been 
seen among Damped Lyman $\alpha$ galaxies (DLAs,
Centuri\' on et al. \cite{miriam}), 
however the number of unmixed giants
with measured nitrogen abundances,  
at extremely
low metallicities is too small to draw any definitive
conclusion.

\section{Oxygen}

In view of the current debate on the solar
oxygen abundance, which ranges
from the ``oxygen starved'' Sun of Asplund
et al. (\cite{Asplund}), with A(O)=8.66, to 
the ``oxygen rich'' Sun of Ayres et al.
(\cite{Tom}), with A(O)=8.84, 
with the intermediate value of A(O)=8.72,
derived from the latest 3D CO$^5$BOLD
simulation of the solar atmosphere
(Steffen \cite{Steffen}), I warn readers to always
check what is the solar oxygen abundance
adopted by any author before comparing
values found in the literature.

For a long time it has been considered
as firmly established that the [O/Fe] ratio 
is constant among metal poor stars at 
a value [O/Fe]$\sim +0.4$, largely relying
on the measurement of the [OI] 630 \, nm line
in metal-poor giants by Barbuy (\cite{barbuy88}).
This view was first challenged by Abia \& Rebolo
(\cite{abia}), who found a linear increase
of [O/Fe] with decreasing metallicities, based
on the measurement of the OI permitted triplet
at 777\, nm.
A similar increasing trend of [O/Fe] was found 
by measurements based on the OH UV lines
(Israelian et al. \cite{israelian98};
Boesgaard et al. \cite{Boesoxy};
Israelian et al. \cite{israelian01}).
On the other hand a  ``plateau'' was found by Carretta et al.
(\cite{carretta2000}), based on the measurement
of both forbidden and permitted lines.

I think everybody now agrees that these discrepancies
are rooted in the limitations of the analysis employed.
In particular the combined effects of deviations from
LTE and the effects of granulation (also referred to, as
``3D effects'') on the different lines
are difficult to assess.

In Fig. 1, I assembled the oxygen measurements
in three recent papers but selecting
{\em only} measurements based on the
[OI] forbidden line, which is the only
line measured at the lowest metallicities, 
to avoid mismatches
due to the use of different indicators. 
Taken at face value these
results seem to support the existence of a linear
increase in [O/Fe], note however that the results
of Cayrel et al. (\cite{cayrel04}) have not been 
corrected for the effects of granulation, because
at present, no 3D model atmospheres suitable for 
this sample of metal-poor
giants exist.
The question remains therefore totally open and awaits
a better understanding of the structures of the atmospheres
of metal-poor stars.

\section{Other $\alpha-$chain elements}

All the even elements from oxygen to titanium can be
formed by successive additions of $\alpha$ particles,
the so-called $\alpha-$chain.
The evolution of all the elements, Ne to Ti, 
should therefore resemble that of oxygen, 
although it need not to be necessarily identical,
since the different elements are made in different
layers, see Limongi \& Chieffi (\cite{limongi})
for a review.

Mg, Si, Ca and Ti all show a plateau
with constant [X/Fe] $\sim +0.4$, according to several
recent investigations (Cayrel et al. \cite{cayrel04};
Cohen et al. \cite{cohen04}; Barklem et al.
\cite{barklem05}).
For sulphur, instead, there is a situation
similar to that of oxygen, with claims
of a plateau-like behaviour 
(Chen et al. \cite{chen02,chen03};
Nissen et al. \cite{nissen}; Ryde \& Lambert
\cite{ryde}), or
of a linear rise 
with decreasing
metallicities (Israelian \& Rebolo \cite{israelian};
Takada-Hidai et al. \cite{takeda}).
Korn \& Ryde (\cite{Korn05}) suggested
that this was linked to the use of different \ion{S}{i}
lines by different investigators. However
from the large sample of stars analysed by 
Caffau et al. (\cite{Caffau}), 
who made use of lines of three different 
\ion{S}{i} multiplets,
it is obvious
that at low metallicities some stars
are found with [S/Fe]$\sim +0.4$, while others
are found with considerably higher [S/Fe]
ratios, even when the {\em same } lines are used.
Whether this discrepancy can  be traced back
to inadequacies of the model atmospheres
or wrong atmospheric parameters, or whether
it reflects a real scatter or bimodality
in the Galactic [S/Fe] distribution,  is still
an open question. 

\section{Iron peak elements}

For iron peak elements the trends
among extremely metal-poor stars already
highlighted by McWilliam et al. (\cite{Mcwil})
and Ryan et al. (\cite{RNB}) have been
confirmed by the recent investigations
(Cayrel et al. \cite{cayrel04}; Cohen et al.
\cite{cohen04}; Barklem et al. \cite{barklem05}).
Namely Cr and Mn show an increasing deficiency
over iron,  with decreasing metallicity. Co, on
the contrary, displays an increasing overabundance,
while Ni tracks iron down to the lowest metallicities.
The new important result of the recent investigations
is the tiny star to star scatter in these
abundance ratios at any given metallicity, showing that
the scatter seen in the early investigations was entirely
due to observational errors.
Copper has been traced down to a metallicity
of [Fe/H] $\sim -3$ using the \ion{Cu}{i} 327\, nm line
(Bihain et al. \cite{bihain}) and the [Cu/Fe] ratio shows a marked 
decrease from the solar value,
over the range $-2\la{\rm [Fe/H]}\la -1$.
Below [Fe/H]$\sim -2$ it has a plateau-like 
behaviour with [Cu/Fe]$\sim -1$. 
Zinc tracks iron ([Z/Fe]=0) down to 
[Fe/H]=--2 (Bihain et al. \cite{bihain})
while at lower metallicities a linear increase
is seen in [Zn/Fe] up to [Zn/Fe]$\sim +0.7 $
at metallicities [Fe/H]$\sim -4$ (Cayrel et al. \cite{cayrel04}).
It is remarkable that, in the metallicity range over
which Zn tracks Fe, the ratio [S/Zn] correlates 
very well [$\alpha$/Fe] (Caffau et al. \cite{Caffau}),
which justifies the use of [S/Zn] as a ``dust--free'' proxy for
[$\alpha$/Fe] in DLAs (Centuri\' on et al. \cite{miriamszn})
and Blue Compact Galaxies.

\section{Odd Z elements: Na, Al, K, Sc}

The production of these elements is sensitive to the neutron
excess (Arnett \cite{arnett}) and may be influenced also by
proton capture reactions
occurring in the red-giant phase through the Ne-Na cycle 
(Langer et al. \cite{langer}), and the Mg-Al cycle
(Denissenkov \& Tout \cite{denissenkov}).
For Na the results of Cayrel et al. (\cite{cayrel04})
indicate a steady overdeficiency of Na over iron with
decreasing metallicity. However these results are based
on the \ion{Na}{i} D lines, which are sensitive to departures
from local thermodynamic equilibrium.
Detailed NLTE computations by Andrievsky et al. (\cite{sergei})
show instead,
that over the range $-4 \le {\rm[Fe/H]}\le -2.5$
there is a constant overdeficiency of sodium over
both iron and magnesium, with
[Na/Fe]$\sim -0.21$ and [Na/Mg]$\sim -0.45$, for both 
``unmixed'' giants and dwarfs.
A few of the observed ``mixed'' giants display overabundances
of sodium, which are likely the result of deep mixing
and operation of the Ne-Na cycle.
The combined effect of departures from LTE and 
granulation effects still needs to be determined.
For Al the data indicate also a constant
value with [Al/Fe]$\sim 0$, with a few Al enhanced
stars among the ``mixed'' giants.
Potassium has been observed for the first time in a large
sample of stars by Cayrel et al. (\cite{cayrel04}),
both K and Sc seem to show a very gentle decrease
in their [X/Fe] ratios with a moderate scatter
of $\sim 0.12$ dex, always close to [X/Fe]=0,
although this slope is not very significant and a larger
number of stars is needed to confirm its reality.

\section{Neutron capture elements}

The most obvious observational feature is the 
large scatter in the [X/Fe] ratios, where X is any
element with Z$> 30$, compared to the tiny scatter
observed for the lighter elements. There is general 
consensus on the fact that this scatter is intrinsic
and not driven by observational errors.
In spite of the scatter
one may notice trends of decreasing [Sr/Fe], [Y/Fe], [Ba/Fe]
and increasing [Eu/Fe] with decreasing metallicity.
Johnson \& Bolte (\cite{jennifer})
have been able to measure Pd and Ag
in three very metal-poor stars, too few
to be able to establish trends, however
the [Pd/Ag] ratios in these stars show 
a real difference from the 
corresponding solar system $r-$process ratio.

One exciting discovery has been the existence
of the so-called $r-$enhanced stars, in which 
$r-$process products are enhanced, sometimes 
by a factor of ten or more, over iron.
The first such star star discovered
CS 22892-052 (Sneden et al. \cite{sneden})
appeared to be unique, however, the discovery
of CS 31082-001 (Cayrel et al. \cite{cayrel01}),
for which the abundance of uranium has been
measured (Cayrel et al. (\cite{cayrel01}; Hill et al. 
\cite{hill}) convinced people that 
there should be many stars of this class.
In fact the dedicated survey 
HERES  (Christlieb et al. \cite{heres},
Barklem et al. \cite{barklem05}) discovered
41 new ones and was furthermore
able to demonstrate that 
they can be conveniently divided into two classes,
$r-$II stars, for which [Eu/Fe] $> 1.0$  and 
$r-$I stars, for which $0.3\le [{\rm Eu/Fe}] \le 1.0$.
Quite interestingly
the $r-$II stars, are 
found in  a narrow metallicity range centred at 
[Fe/H] $\sim  -2.8$, 
with a tiny scatter (0.16 dex). 
The $r-$I stars, on the other hand,  
are found across the entire metallicity range 
covered by the HERES survey and are much more
common.
Recently Frebel and collaborators
(Frebel et al. \cite{anna}) have 
discovered another $r-$II star, HE 1523-0901
with  [Fe/H] = --3.1 and [Eu/Fe] = +1.8.

A very intriguing result of the study of the
$r-$enhanced stars is that there are probably 
multiple sites for the $r-$process, which
would thus not be ``universal'' (Hill et al. \cite{hill};
Ishimaru et al. \cite{yuhri}; Wanajo \& Ishimaru
\cite{WI}; Honda et al. \cite{honda}).
An example of our still poor understanding
of the $r-$process is the measurement
of lead in CS 31082-001 by Plez et al. (\cite{plez}),
the amount of lead measured is essentially 
all which is expected from the decay of Th and U, 
leaving none for the $r-$process production. 

Lead can also be formed in the $s-$process and more
efficiently in very metal-poor stars, due to the high
neutron to iron seed ratio.
The discovery of lead-rich
extremely metal poor stars,
like CS 29497-030 with [Pb/Fe]=+3.5 (Sivarani et al.
\cite{sivarani}), demonstrates that the $s-$process
is operating also at extremely low metallicities.

\section{The hyper-metal-poor stars}

This class of stars defined  by Beers \& Christlieb
(\cite{BC}) as stars with 
$-5< {\rm [Fe/H]}\le -6$ currently
contains  only two stars: HE0107-5240 (Christlieb et al.
\cite{christlieb}) and HE1327-2326 (Frebel et al.
\cite{frebel}).
It should be however noted that these two stars have
an extraordinary chemical composition with large
enhancements of C,N and O so that their
metallicity, meant as $Z$, is, in spite
of the name of the class, comparable to
that of metal-poor globular cluster stars.
The recent downward revision of C,N,O
abundances in these stars, based on 3D model atmospheres
and LTE line formation,
by Collet et al. (\cite{collet}), 
places these stars at considerably lower $Z$.

\section{CEMP stars}

The Carbon Enhanced Metal Poor (CEMP) stars,
are stars for which carbon is enhanced over
iron. Beers \& Christlieb (\cite{BC}) suggest
to define the class as stars with [C/Fe]$>+1$
although there is no general agreement on this.
Among these some are true ``carbon stars'', i.e.
C/O$ >1$, spectroscopically recognisable by the presence
of strong C$_2$ Swan bands.
There is an ongoing debate on the frequency of CEMP
stars, estimated to be $\sim 25$\% by
Marsteller et al. (\cite{marsteller}) 
and only 14\% by Cohen et al. (\cite{jud05}).
Many of these stars are binary and the C-enhancement
is the result of mass-transfer from an AGB
companion, currently in the white dwarf stage.
Lucatello et al. (\cite{lucatello}) found
that the detected binary 
fraction among CEMP stars {\em with } enhancement
of $s-$process elements is 68\%, implying that
the binary fraction among these stars is higher than 
among field stars. These authors also contend that
in fact {\em all} these stars are in double
or multiple systems.

It is quite likely that the CEMP class
includes very diverse objects,
Beers \& Christlieb (\cite{BC})
propose
the following sub-classes:
 CEMP-r, which exhibit $r-$process enhancement,
the prototype star is  CS 22892-052 (Sneden et al. \cite{sneden});
CEMP-s, which exhibit 
$s-$process enhancement, prototype stars
LP 625-44 and LP 706-7 (Norris et al. (\cite{NRB});
CEMP-r/s, which exhibit an enhancement pattern that can be attributed 
to $s-$process {\em and} $r-$process,
prototype star HE 2148-1247 (Cohen et al. \cite{jud03});
CEMP-no, stars which exhibit
no enhancement of neutron capture elements, 
prototype star CS 22957-027 (Norris et al. \cite{N97};
Bonifacio et al. \cite{B98}).
Recently Sivarani et al. (\cite{siva06})
have discovered 2 CEMP dwarf stars CS 29528-041 and 
CS 31080-095 which show no enhancement in Sr, but
a moderate enhancement in Ba, for which they suggest
a new class of CEMP-no/s which could constitute 
a link between the CEMP-no and the CEMP-s stars.

\begin{acknowledgements}

I am grateful to the organisers of this meeting
for inviting me to write this review.
I wish to thank Anna Frebel for providing me 
information on the newly discovered $r-$II star
HE 1523-0901 in advance of publication.
I acknowledge support 
from EU contract
MEXT-CT-2004-014265 (CIFIST).

\end{acknowledgements}

%%-----------------------------
%%      your bibliography
%%-----------------------------

\end{document}